\documentclass[12pt,a4]{article}
\usepackage{latexsym}
\usepackage{amsmath}
\usepackage{bm}
\usepackage{graphicx,color}
\usepackage{txfonts}
\newcommand{\bea}{\begin{eqnarray}}
\newcommand{\eea}{\end{eqnarray}}

\newcommand{\la}{\label}

\title{Young's Double Slit Experiment in Quantum Field Theory }
\author{Masakatsu KENMOKU\thanks{kenmoku@asuka.phys.nara-wu.ac.jp} 
\ and Kenji KUME\thanks{kume@cc.nara-wu.ac.jp} \\
Department of Physics, 
Nara Women's University, Nara 630-8506, Japan
}
\date{\empty}

\begin{document}
 %====================================================
\maketitle
\abstract{
Young's double slit experiment is formulated in 
the framework of canonical quantum field theory  
in view of the modern quantum optics. 
We adopt quantum scalar fields instead of quantum electromagnetic fields 
ignoring the vector freedom in gauge theory.   
The double slit state is introduced in Fock space 
corresponding to experimental setup. 
As observables, expectation values of energy density 
and positive frequency part of current 
with respect to the double slit state are calculated which 
give the interference term.  
Classical wave states are realized 
by coherent double slit states in Fock space 
which connect quantum particle states 
with classical wave states systematically. 
In case of incoherent sources, the interference term vanishes 
by averaging random phase angles as expected. 
}

%------------------------------------------
\section{Introduction}
%-------------------------------------------
\renewcommand{\theequation}{\thesection.\arabic{equation}}
\setcounter{equation}{0}
In the early nineteen century, Thomas Young performed 
his famous double slit diffraction experiment using a light source, 
which shows the interference pattern on the screen 
\cite{BornSmith}.  
The interference effect is explained 
in the framework of classical optics 
by the method of 
Huygens-Fresnel principle in classical optics theory {\cite{Rossi}}. 
As a modern version of this experiment, 
the quantum double slit experiments are performed 
using photons, electrons, neutrons and others. 
It is necessary to use quantum mechanics 
in order to explain the interference pattern 
for sub-atomic particles {\cite{Feynman}}. 

\begin{figure}[h]
\begin{center}
\includegraphics[width=8cm]{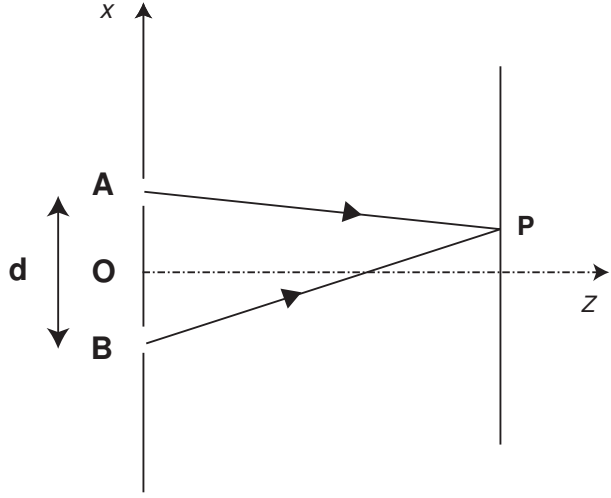}
\end{center}
\caption{Schematic design of Young's double slit experiment 
between slits and screen.
Coordinates are assigned as the origin $O=(0,0,0)$, slits  
A=(d/2,0,0), B=(-d/2,0,0) and observation position $P=(x,y,z)$.}
\end{figure}

%We study the double slit experiment 
%which includes the double-slit experiment and takes account of 
%the incoherent sources as well as coherent sources \cite{Paul}.
In standard quantum mechanics, 
the double slit wave function is 
obtained as a special solution of 
the Shr${\ddot{{\rm o}}}$dinger wave equation 
as the superposition of waves emitted from slits A, B   
at the points $(d/2,0,0)$, $(-d/2,0,0)$ 
\bea
{F}^{DS}_k({\bm r},t)
={F}_{k}^A({\bm r},t)+{F}_{k}^B({\bm r},t) 
\ ,\label{eq:101}\eea
where each solution is 
expressed by spherical waves as a good approximation 
in the region between slits and screen:   
\bea
 {F}^{A}_{k}({\bm r},t)
=A_{0}\,{{\exp}({-i\omega_{k} t+ikr_{A}})}/{r_{A}} \ , \ 
{F}^{B}_{k}(r,t)
=B_{0}\,{{\exp}({-i\omega_{k} t+ikr_{B}})}/{r_{B}}
\ ,\label{eq:102}\eea
where 
$\bm{r}=(x,y,z)$ denotes the observation point on the screen and 
$r_{A},r_{B}$ are distances between each slit and the observation point:
\bea 
r_{A}=\sqrt{(x-d/2)^2+y^2+z^2}\ ,\ r_{B}=\sqrt{(x+d/2)^2+y^2+z^2}
\ .\eea
The wave number 
and frequency are denoted by 
$k=(k_{x},k_{y},k_{z})$ with $k=|{\bm k}|$ 
and $\omega_{k}$.
Amplitudes include fixed absolute values and phases for double-sources:
$A_{0}=|A_{0}|\exp{}i\theta_{A},B_{0}=|B_{0}|\exp{i\theta_{B}}$.  
Intensity of wave function is given as  
\bea
|{F}^{DS}_{k}(\bm{r},t)|^2
&=&|{F}^{A}_{k}(r,t)|^2+|{F}^{B}_{k}(r,t)|^2
+2\,{\rm Re}({{F}^{A}_{k}(r,t)^{\ast}{F}^{B}_{k}(r,t)})
\ ,\eea
where the interference effect appears in the last term as
\bea
2{\rm Re}({{F}^{A}_{k}(r,t)^* {F}^{B}_{k}(r,t))}
&=&2\frac{|A_{0}||B_{0}|}{r_{A}r_{B}}
\cos{(k(r_{A}-r_{B})+\theta_{A}-\theta_{B})}
\nonumber\\
&\simeq& 2\frac{|A_{0}||B_{0}|}{r^2}\cos (-kxd/r+\theta_{A}-\theta_{B})
\ ,\label{eq:105}
\eea
which holds for $d \ll r=\sqrt{x^2+y^2+z^2}$.  
The intensity is interpreted as the probability density 
in the non-relativistic Shr${\ddot {\rm o}}$dinger theory
because it is the density of the conserved current.   
We introduce each phase for each source in order to 
treat incoherent cases as well as coherent cases. 
We are able to obtain the same interference term for the classical 
wave mechanics including classical electromagnetism as the 
non-relativistic quantum mechanics.  

Photons and particles in high speed, however, should be treated 
in the framework of quantum theory with relativistic covariance.  
The relativistic quantum mechanics is not complete theory 
because of some difficulties: 
the existence of negative-energy solutions, 
the luck of probability interpretation in Klein-Gordon theory,  
the non-existence of charge conjugation states in Dirac theory
\cite{Sakurai}. 
Therefore relativistic particles should be treated 
as the relativistic quantum field theory  
taking account of the particle creation and annihilation 
{\cite{fieldtheory}}. 

Several works have been done in this direction, where 
radiation waves from sources A and B are described by 
two spherical waves in quantum field theoretical treatment  
but on the other hand those are described by plane-waves 
in classical treatment {\cite{Scully,Rand}} .    
In this paper, we study the method that the  
radiation waves from two sources are treated as two spherical waves 
in quantum field theory. 
Our method shows that the spherical wave states, A and B, 
are not independent $<A|B>\neq 0$  
different from the plane-wave treatment as $<A|B>=0$,  
and can naturally connect the description of Young's double slit  
experiment in quantum field theory with that in quantum mechanics 
and in classical wave mechanics.  
{\footnote{
The plane-wave treatment can reproduce the same interference effect 
as the classical wave mechanics but cannot express the 
global structure of space as the distance dependence 
on sources and the screen.}}
  
We formulate the Young's double slit experiment 
in standard canonical formalism of quantum field theory   
by introducing the double slit states in Fock space. 
Our basic idea is that quantum fields are treated in 
general canonical form and state vectors are 
in a special form corresponding to the 
experimental setup or boundary condition  
by introducing the double slit states in Fock space.  
We formulate it in the quantum scalar field theory 
instead of the quantum electrodynamics  
avoiding the complexity arising from 
consistent compatibility of gauge invariance with 
relativistic covariance. 
We are able to understand the relation in the interference effects    
among classical wave mechanics, quantum mechanics and quantum field
theory by studying the quantum field theory 
%ignoring photon helicity freedom 
through the coherent state method {\cite{Glauber}}. 

The organization of this paper is the following. 
In section 2, the general description is shown  
for the Young's double slit experiment 
in the relativistic quantum field theory 
for massive and/or massless scalar field. 
In subsection 2.1 
the canonical quantization for quantum scalar field is reviewed  
for the sake of following study. 
In subsection 2.2  general formalism for double-source states 
is introduced in Fock space. 
In subsection 2.3 application of the general formalism 
to the Young's double slit experiment is shown.    
In section 3 classical waves are realized 
by introducing the coherent double slit states in Fock space. 
In section 4 case of incoherent sources are studied. 
In the final section, the result is summarized and some discussions 
are given.   

\section{Double slit experiment in scalar field theory}
\setcounter{equation}{0}
Field theoretical approach to the double slit problem is formulated 
for field operators as general solutions of operator field equations  
and for state vectors in Fock space connecting special solutions 
of c-number field equations.  
The Heisenberg picture is taken in the following.

%%%%%%%%%%%%%%%%%%%%%%%%%%%%%%%%%%%%%%%%%%%%%%%%%%%%%%%%%%%%%%%%%%%%%%%%
\subsection{Canonical scalar field theory}
%%%%%%%%%%%%%%%%%%%%%%%%%%%%%%%%%%%%%%%%%%%%%%%%%%%%%%%%%%%%%%%%%%%%%%%%

In this subsection, we review and summarize quantum field theory 
for the following convenience. 
Here we take quantum free scalar theory  
because we mainly consider their propagation in free space  
between slits and screen. We also ignore 
the helicity freedom of electromagnetic fields 
because we focus on the interference effects not special to each 
helicity state. 

We start from the action and the Lagrangian for 
free scalar quantum field theory in the natural unit: $c=\hbar=1$ as 
\bea
\displaystyle
I&=&\int dt L(t) \ ,\\
L(t)&=&\int d^3x (({\dot\Phi}({\bm r},t)^2
-(\nabla{\Phi(\bm r , t))^2}
-\mu^2\Phi({\bm r},t)^2)/2\ , 
\eea 
where $\dot \Phi$ and $\mu$ denote the time derivative 
and the mass term for the scalar particle. 
The canonical momentum with respect to the scalar field 
is defined and obtained:
\bea
\Pi({\bm r},t)&=&\delta L(t)/\delta {\dot\Phi({\bm r},t)}
=\dot\Phi({\bm r},t)\ ,
\eea
and the canonical equal-time commutation relations are imposed:
\bea
[\Phi({\bm r},t), \Pi({\bm r'},t)]=i\delta^{(3)}({\bm r}-{\bm r'})\ , 
[\Phi({\bm r},t), \Phi({\bm r'},t)]=[\Pi({\bm r},t), \Pi({\bm r'},t)]=0
\ .\label{eq:204}\eea
The Hamiltonian and Hamiltonian density are defined as 
\bea
H&=&\int d^3 x \Pi({\bm r},t)\dot\Phi({\bm r},t)-L
= \int d^3 x \,{\cal{H}}({\bm r},t) \ , \\ 
{\cal H}({\bm r},t)
&=& : ({\Pi}({\bm r},t)^2 
+(\nabla{\Phi({\bm r},t))^2}
+\mu^2\Phi({\bm r},t)^2):/2\ , 
\label{eq:206}\eea 
where the mark $:\ :$ denotes the normal order product. 
Using the Heisenberg equation of motion for $\Phi({\bm r},t)$ and 
$\Pi({\bm r},t)$, the field equation for quantum fields is obtained:
\bea 
\ddot{\Phi}({\bm r},t)-\nabla^2\Phi({\bm r},t)-\mu^2\Phi({\bm r},t)=0 
\ .\label{eq:207}\eea
The general solution of the operator field equation is given by 
the expansion of eigen-value solutions 
with operator coefficients $a_{\bm k }$ and $a_{\bm k}^{\dagger}$:   
\bea
\displaystyle
\Phi({\bm r},t)=\sum_{\bm k}  
(u_{\bm k}({\bm r},t)a_{\bm k}+u_{\bm k}({\bm r},t)^{*} a_{\bm k}^{\dagger}) 
\ ,\label{eq:209}\eea
where the plane wave solutions are taken as eigen-value solutions 
with the periodic boundary condition:  
\bea
u_{\bm k}({\bm r},t)=n_{\bm k}\exp{(-i\omega_{k}t+i{\bm k}\cdot{\bm r})} 
\ ,\label{eq:209}\eea
where the wave number, frequency and the normalization factor are 
denoted as ${\bm k}=(k_{x},k_{y},k_{z})$ with $k=|{\bm k}|$, 
$\omega=\sqrt{k^2+\mu^2}$ and 
$n_{k}=1/\sqrt{2\omega_{k}V}$ ($V$: the volume of space) 
respectively. 
They satisfy the ortho-normal relations: 
\bea
(u_{\bm k},u_{\bm {\ell}})
=-(u_{\bm k}^{*},u_{\bm {\ell}}^{*})=\delta_{\bm k,{\ell}}\ , \ 
(u_{\bm k}^{*},u_{\bm {\ell}})
=(u_{\bm k}^{*},u_{\bm {\ell}}^{*})=0\ ,  
\label{eq:210}\eea
where the relativistic inner product is defined:
\bea
(A,B)=i\int d^3x(A^{*}{\dot B}-{\dot A}^{*}B)\ ,
\la{eq:211}\eea
with the completeness relation:
\footnote{
In the completeness relation, 
prescription 
$i{\bm k}\cdot{\bm (r-r')}
\rightarrow i{\bm k}\cdot{\bm(r-r')}-\epsilon|{\bm k}|$ 
is to be understood, with the small and positive regularization 
parameter $\epsilon$.
} 
\bea
i\sum_{\bm k}(u_{\bm k}^{*}({\bm r},t){\dot u}_{\bm k}({\bm r'},t)
-{\dot u}_{\bm k}^{*}({\bm r},t)u_{\bm k}({\bm r'},t))
=\delta^{(3)}({\bm r}-{\bm r'})
\ .\eea

The commutation relations among $a_{\bm k}, {a_{\bm k}}^{\dagger}$ 
are obtained from the canonical commutation relations in eq.(\ref{eq:204}) 
and the ortho-normal relations in eq.(\ref{eq:210}) as
\bea
[a_{\bm k}, {a_{\bm {\ell}}}^{\dagger}]=\delta_{{\bm k},{\bm {\ell}}}\ , \ 
[a_{\bm k}, {a_{\bm {\ell}}}]
=[a_{\bm k}^{\dagger}, {a_{\bm {\ell}}}^{\dagger}]=0
\ .\label{eq:213}\eea
The Hamiltonian is expressed by 
the creation and annihilation operators omitting the zero point energy:
\bea
H=\sum_{\bm k}\omega_{k}a_{\bm k}^{\dagger}a_{\bm k}\ .
\eea

%%%%%%%%%%%%%%%%%%%%%%%%%%%%%%%%%%%%%%%%%%%%%%%%%%%%%%%%%%%%
\subsection{General formalism for state vector}
%%%%%%%%%%%%%%%%%%%%%%%%%%%%%%%%%%%%%%%%%%%%%%%%%%%%%%%%%%%%

In this subsection, we develop the method to form the 
state vector in the Fock space for the general interference experiment 
in the Heisenberg picture. Our method can connect solutions of the quantum 
field theory with those of quantum mechanics and of classical wave mechanics.  

We start to consider a state $|X>$ which is 
the superposition of one particle state in the 
Fock space:
\bea
|X>=\sum_{\bm\ell}f_{\bm\ell}a_{\bm\ell }^{\dagger}\mid 0>\ ,
\la{eq:215}\eea
with the superposition coefficients $f_{\bm\ell}$.  
In order to determine the superposition coefficients $f_{\bm\ell}$, 
the c-number function ${F}({\bm r},t)$ 
of one particle expectation value is introduced: 
\bea
{F}({\bm r},t)=<0|\Phi{({\bm r},t)}\mid X> 
=\sum_{\bm\ell}f_{\bm\ell}u_{\bm\ell}\ , 
\eea 
where $u_{\bm\ell}$ is the eigen-value solution of eq.(\ref{eq:209}). 
\footnote{
If the normalization condition is imposed on the state $<X|X>=1$, 
the superposition coefficients satisfy the condition 
$\sum_{\bm\ell}f_{\bm\ell}^*f_{\bm\ell}=1$. 
Under this condition, the one particle expectation value is also
normalized $({F}({\bm r},t),{F}({\bm r},t))=1$.   
}
As the quantum field $\Phi{({\bm r},t)}$ is the general solution 
of filed equation eq.(\ref{eq:207}), the function ${F}({\bm r},t)$ should satisfy 
the c-number field equation in the same form: 
\bea 
\ddot{F}({\bm r},t)-\nabla^2{F}({\bm r},t)
-\mu^2{F}({\bm r},t)=0 \ . 
\la{eq:217}\eea  
It should be noted that some source terms and/or boundary terms 
can be taken into account in this c-number field 
equation corresponding to the experimental setup. 
If the special solution ${F}({\bm r},t)$ is obtained, 
the superposition coefficients $f_{\bm\ell}$ are obtained as 
the Fourier transformation coefficients:
\bea
f_{\bm\ell}=(u_{\bm\ell}({\bm r},t),{F}({\bm r},t))|_{t=0}, 
\la{eq:218}\eea  
where the round bracket $(A,B)$ denotes the relativistic inner product 
defined in eq.(\ref{eq:211}) 
and thus the corresponding state vector in the Fock 
space is determined as in eq.(\ref{eq:215}). 
In the end of the calculation in eq.(\ref{eq:218}), the time is set to zero 
because the Heisenberg picture is taken. 

%%%%%%%%%%%%%%%%%%%%%%%%%%%%%%%%%%%%%%%%%%%%%%%%%%%%%%%%%%%%%%%%%%%%%%%%%%
\subsection{Double slit state}
%%%%%%%%%%%%%%%%%%%%%%%%%%%%%%%%%%%%%%%%%%%%%%%%%%%%%%%%%%%%%%%%%%%%%%%%%%

As an important application of this general method to form the 
special state in the Fock space, we study the state vectors for 
the Young's double slit experiment.   

In order to obtain the double slit state $|{DS};{k}>$, 
firstly we solve the one particle c-number field equation of 
eq.({\ref{eq:217}}) and the spherical wave solution 
${F}^{DS}_k({\bm r},t)={F}^{A}_k({\bm r},t)+{F}^{B}_k({\bm r},t)$ is obtained 
as the special solution as in  
the non-relativistic quantum mechanics of eqs.(\ref{eq:101}) 
and (\ref{eq:102}).   
The superposition coefficients $f^{DS}_{{\bm\ell},k}$ 
are obtained as the Fourier expansion coefficients of the c-number  
double-source function ${F}^{DS}_k({\bm r},t)$:
\footnote{
In driving the Fourier expansion coefficient, 
prescription $k\rightarrow k+i\epsilon$ is to be understood. 
\label{foot:03}}
\bea
f^{DS}_{{\bm\ell},k}&=&f^{A}_{{\bm\ell},k}+f^{B}_{{\bm\ell},k}\ ,\nonumber\\
f^{A}_{{\bm\ell},k}&=&(u_{\bm\ell},{F}^{A}_{k}/\sqrt{2\omega_{k}})|_{t=0}
=\frac{4\pi n_{\ell}(\omega_{\ell}+\omega_{k})
}{\sqrt{2\omega_{k}}(\ell^2-k^2)}A_{0}{\exp}(-i\ell_{x}d/2)\ , \nonumber\\
f^{B}_{\ell,k}&=&(u_{\bm \ell},{F}^{B}_{k}/\sqrt{2\omega_{k}})|_{t=0}
=\frac{4\pi n_{\ell}(\omega_{\ell}+\omega_{k})
}{\sqrt{2\omega_{k}}(\ell^2-k^2)}B_{0}{\exp}(i\ell_{x}d/2)
\ , \la{eq:220}\eea
where ${\ell}=(\ell_{x},\ell_{y},\ell_{z})$ is the linear momentum
and $n_{\ell}$ is the normalization factor of the plane waves 
given below the eq.(\ref{eq:209}). 
At the end of calculation, the time variable 
for the double slit state is set to be
zero because the Heisenberg picture is taken. 
Additional factor $\sqrt{2\omega_k}$ is included for the 
normalization of relativistic inner product. 
The double-source state 
$|{DS};{k}>$ for a fixed wave number $k=|{\bm k}|$ 
is formed in Fock space 
with the superposition coefficients $f_{{\bm\ell}, k }$  
: 
\bea
\displaystyle
|{DS};{k}>=\sum_{\bm\ell}a_{\bm\ell}^{\dagger}|0>f^{DS}_{{\bm\ell},k}
\ .\la{eq:219}\eea
We can confirm that one particle expectation value 
on the screen position ${\bm r}$   
\bea
<0|\Phi({\bm r},t)|{DS};{k}> 
=
({F}^{A}_{k}({\bm r},t)+{F}^{B}_{k}({\bm r},t))/\sqrt{2\omega_{k}}
={F}^{DS}_{k}({\bm r},t)/\sqrt{2\omega_{k}}
\ ,\la{eq:221}\eea
reproduces the same spherical wave function in 
quantum mechanics or classical wave mechanics    
for the double-slit experiment. 
\footnote{
Of course, the frequency $\omega_{k}$ is $\sqrt{k^2+\mu^2}$ for
relativistic theory and $k^2/2\mu$ for non-relativistic theory.} 

Note that the one-particle expectation value 
$<0|\Phi({\bm r},t)|{DS};k> $ in eq.(\ref{eq:221}) 
does not have the meaning of 
provability amplitude of wave function because of the 
non-existence of the consistent relativistic quantum mechanics 
but plays the role of provability density for conserved quantities. 
For the case that the positive $\Phi^{(+)}({\bm r},t)$
and negative $\Phi^{(-)}({\bm r},t)$ 
frequency parts of the scalar fields can be separated definitely, 
which is possible for the free field theory, the 
conserved current for $\Phi^{(+)}({\bm r},t)$ can be constructed 
\cite{Bjorken}:  
\bea
j^{(+)}_{\mu}({\bm r},t)=
-i(\Phi^{(+)}({\bm r},t)^{\dagger}\partial_{\mu}\Phi^{(+)}({\bm r},t)-
\partial_{\mu}\Phi^{(+)}({\bm r},t)^{\dagger}\Phi^{(+)}({\bm r},t)) \ , 
\eea
where
\bea
\Phi^{(+)}({\bm r},t)=\sum_{k}u_{k}({\bm r},t)a_{k}\ .
\eea
Its density is a candidate for good physical
observables and the expectation value for the double slit state is 
calculated as  
\bea
<{DS};k|j^{(+)}_{0}({\bm r},t)|{DS};k>
=|{F}^{DS}_{k}(\bm r, t)|^2 \ ,
\la{eq:224}\eea 
which is the expected result. 
Another good candidate for observables is the energy density  
because the total energy conserves. 
The expectation value of energy density on the screen position ${\bm r}$ 
with respect to the double-source state is given as 
\footnote{
The higher order term $O(1/r^3)$ 
comes from the calculation of the term $(\nabla \Phi(r,t))^2/2$ 
in eq.(\ref{eq:206}). 
} 
\bea
<{DS};k|{\cal H}({\bm r},t)|{DS};k>
= \omega_{k}|{F}^{DS}_{k}({\bm r},t)|^2\ +O(1/r^3)\ ,
\la{eq:225}\eea 
which is again the expected form for the double slit experiment. 

%%%%%%%%%%%%%%%%%%%%%%%%%%%%%
It should be noted that the density matrix method for the 
quantum observable is applied to our method as
\bea
<DS;k|{\cal O}|DS;k>=Tr \{ \rho^{DS}_{k} {\cal O}\} \ \ \  
{\mbox{for}}\ \ \ 
 {\cal O}=j^{(+)}_{0}({\bm r},t) \ , \  {\cal H}({\bm r},t)
\ ,\eea 
where the density matrix 
is $\rho^{DS}_{k}=|DS;k><DS;k|$.
{\footnote{The double-source state $|DS;k>$ is not a normalized state
in general but can be a normalized one: $|DS;k>/\sqrt{<DS;k|DS;k>}$.
}} 
%%%%%%%%%%%%%%%%%%%%%%%%%%%%%%%%%%%%%%
 
We have obtained the same interference pattern 
for the famous Young's double slit experiment 
in the quantum field theoretical description 
of eqs.(\ref{eq:224}) and (\ref{eq:225}) with the 
quantum mechanical description of eq.(\ref{eq:105}) 
in our double slit state method given in eq.(\ref{eq:219}). 

%%%%%%%%%%%%%%%%%%%%%%%%%%%%%%%%%%%%%%%%%%%%%%%%%%%%%%%%%%%%%%%%%
\section{Classical waves as coherent double-source state}
%%%%%%%%%%%%%%%%%%%%%%%%%%%%%%%%%%%%%%%%%%%%%%%%%%%%%%%%%%%%%%%%%

\setcounter{equation}{0}

In this section, 
we are going to study the relation between 
the relativistic quantum field theory and the corresponding classical 
wave mechanics for the interference experiment. 

In order to establish the relation, 
we introduce the general coherent state  
$|{CX}>$ for the c-number wave function ${F}({\bm r},t)$
as the eigen-value function of the positive frequency 
part of the general operator field $\Phi^{(+)}({\bm r},t)$:
\bea
\Phi^{(+)}({\bm r},t)|CX>={F}({\bm r},t)|CX> \ .
\la{eq:301}\eea
As the positive frequency part of quantum field $\Phi^{(+)}({\bm r},t)$ 
satisfies the field equation in eq.(2.7), 
the eigen-value function should satisfy the same field equation. 
Because the positive frequency part contains only annihilation operators 
$a_{\bm \ell}$, the eigen-state $|{CX}>$ is given by 
the exponential function of the superposition 
of creation operators $a_{\bm \ell}^{\dagger}$, that is the coherent state, as
\bea
|{CX}>=\exp{(\sum_{\bm\ell}f_{\bm\ell}a_{\bm\ell }^{\dagger})} \mid 0>\ .
\la{eq:302}\eea
Using commutation relations in eq.(\ref{eq:213}), 
the following relations are derived from eqs.(\ref{eq:302}) 
and (\ref{eq:301}) as      
\bea
a_{\bm\ell}|CX>=f_{\bm\ell}|CX>\ \ ,\ \ 
\sum_{\bm\ell}f_{\bm\ell}u_{\bm\ell}
={F}({\bm r},t)\ . 
\eea 
Then the superposition coefficient $f_{\bm\ell}$ is again given by the 
Fourier transformation coefficient of ${F}({\bm r},t)$ given in
eq.(\ref{eq:218}). The method establish to make the relation of 
the general quantum filed to special c-number field. 
 
This general method is applied to the case of double slit experiment. 
The coherent double slit state $|{CDS};{k}>$ 
is obtained as the coherent state according to the above method as 
\bea
|{CDS};{k}>
=\exp{(\sum_{\bm\ell}a_{\bm\ell}^{\dagger}f^{DS}_{{\bm\ell},k})}|0>\ ,
\eea
which satisfies the eigen-value equation: 
\bea
\Phi^{(+)}({\bm r},t)|{CDS};{k}>={F}^{DS}_{k}({\bm r},t)
|{CDS};{k}> \ ,
\eea
where  
the superposition coefficients are given as the same forms 
 in eq.(\ref{eq:220}). 
This coherent double slit state is the extended application 
of the coherent state for the monochromatic state 
$\exp{\alpha a^{\dagger}}$, 
which satisfies $a\exp{\alpha a^{\dagger}}=\alpha \exp{\alpha a^{\dagger}}$, 
to the spherical waves with definite wave
length. 

One-particle expectation value 
for the coherent double slit state is given as 
\bea
<0|\Phi({\bm r},t)|CDS;{k}> 
=(F^{A}_{k}({\bm r},t)+F^{B}_{k}({\bm r},t))/\sqrt{2\omega_{k}}
=F^{DS}_{k}({\bm r},t)/\sqrt{2\omega_{k}}
\ , \eea 
which is the same as the one for the double slit state 
in eq.(2.21).
The expectation value 
of the Hamiltonian density for the coherent double slit state is 
obtained as 
\bea
<CDS;{k}|{\cal H}({\bm r})|CDS;{k}>/<CDS;{\bm k}|CDS;{k}>
= \omega_{k}|F^{DS}_{k}({\bm r},t)|^2\ +O(1/r^3)\ 
,\eea  
which is also the same as the one for the double slit state 
in eq.(\ref{eq:225}).  

The coherent double slit state is 
the set of the Poisson distribution for quantum photons  
and the operation of the annihilation operators to this state 
makes them to classical Fourier expansion coefficients 
which correspond to the classical spherical waves. 
The coherent double slit state method connects  
classical electromagnetic wave states with 
quantum photon states, because 
the coherent double slit state can reduce to one photon state 
by choosing the values of amplitudes 
$A_{0},B_{0}$ small (see eq.(\ref{eq:220})).  

It should be noted that 
the coherent double slit  state method 
for spherical waves is more general than the  
usual coherent state method for monochromatic plane waves like laser
beams, which is produced by the continuous induced emission process 
of photons.
Some general and interesting 
investigations on coherent states have been done 
extensively{\cite{Glauber}}. 

%%%%%%%%%%%%%%%%%%%%%%%%%%%%%%%%%%%%%%%%%%%%%%%%%%%%%%%%%%%%%%%%%%%%%%%%% 
\section{Case of incoherent sources}
%%%%%%%%%%%%%%%%%%%%%%%%%%%%%%%%%%%%%%%%%%%%%%%%%%%%%%%%%%%%%%%%%%%%%%%%%
\setcounter{equation}{0}

We have studied the case of light sources A,B with definite phases 
$\theta_{A},\theta_{B}$ for quantum photons in the double slit state 
and coherent double slit state for classical waves 
in the previous subsections.   
Next we study the case of two incoherent sources. 
In this case the phases for two sources A and B are incoherent 
and random with each other 
and therefore observable values are expected as the average values of them.  
The energy density and the positive frequency part of the current density 
in case of incoherent sources are 
\bea
\int_{0}^{2\pi}\frac{d\theta_{A}}{2\pi}
\int_{0}^{2\pi}\frac{d\theta_{B}}{2\pi}
<DS;{k}|{\cal H}({\bm r})|DS;{k}>
\simeq \omega_{k}(|F^{A}_{k}({\bm r},t)|^2+|F^{B}_{k}({\bm r},t)|^2)\ ,
\nonumber\\
\int_{0}^{2\pi}\frac{d\theta_{A}}{2\pi}
\int_{0}^{2\pi}\frac{d\theta_{B}}{2\pi}
<DS;{k}|j^{(+)}_{0}({\bm r},t)|DS;{k}>
=|F^{A}_{k}({\bm r},t)|^2+|F^{B}_{k}({\bm r},t)|^2\ ,
\eea 
where the interference terms disappear. 
The expectation values for the coherent double slit states 
$|CDS;k>$, 
the interference terms are shown to disappear.   

Note that our method is quite different from the one 
that the double slit state in Fock space is formed as 
the superposition of the independent state of A and B, 
independent monochromatic wave states ${\bm k}_{A},{\bm k}_{B}$ or 
independent spherical wave states. 
Indeed our spherical states A and B are not independent but 
their inner product has non-zero value:
\bea
<B;{k}|A;{k}>\propto B_{0}^*A_{0}\frac{\sin{kd}}{kd}\ ,
\eea    
where each spherical states are defined 
\bea
|A;{k}>:=\sum_{\bm \ell}a_{\bm \ell}^{+}|0>f^{A}_{{\bm\ell},k}\ , \ \  
|B;{k}>:=\sum_{\bm \ell}a_{\bm \ell}^{+}|0>f^{B}_{{\bm\ell},k}\ , 
\eea 
with the Fourier coefficients given in eq.(2.20).
\footnote{
The exact expression of the inner product is 
$<B;{k}|A;{k}>=2\pi B_{0}^*A_{0}{\sin{kd}}/{\epsilon kd}$, 
where $\epsilon$ is the regularization parameter introduced 
in the calculation of Fourier transformation; see footnote {\ref{foot:03}}.}
We can confirm that the limiting cases of the inner product of eq.(4.2) 
coincides the norm itself in short separation limit and tends to zero 
in infinitely separation limit: 
\begin{eqnarray}
<B;{k}|A;{k}>\rightarrow 
\left\{
\begin{array}{cl}
<A;{k}|A;{k}>\ & {\mbox{for}}\ \ d\rightarrow 0 \ {\mbox and}\ B_{0}=A_{0} 
\\
 0 \ & {\mbox{for}}\ \ d\rightarrow \infty  
\end{array}
\right. \ . 
\end{eqnarray}
This result confirms that our method is reasonable and consistent 
for the incoherent sources of interference phenomena.  

%%%%%%%%%%%%%%%%%%%%%%%%%%%%%%%%%%%%%%%%%%%%%%%%%%%%%%%%%%%%%%%%%%%%%%%%
\section{Conclusion and discussions}
%%%%%%%%%%%%%%%%%%%%%%%%%%%%%%%%%%%%%%%%%%%%%%%%%%%%%%%%%%%%%%%%%%%%%%%%

We have formulated the famous Young's double slit experiment in 
quantum field theory based on the modern quantum optics 
by introducing the double slit state in Fock space $|DS;k>$. 
As good physical observables the expectation values of 
energy density ${\cal H}({\bm r},t)$ 
and current density of positive frequency part $j^{(+)}_{0}({\bm r},t)$ 
with respect to the double slit state for quantum photons 
are calculated and the interference terms are obtained as the classical 
Young's experiment. 

The connection of the quantum field theoretical method with 
the classical wave mechanical one is established by 
introducing the coherent double slit state in Fock space $|CDS;k>$, 
which is the double slit eigen-state of the annihilation operator.
The expectation values of physical observables 
with respect to the coherent double slit state 
are calculated and the interference terms are obtained 
as the quantum photon case. 
Our formulation can connect quantum photon state with 
classical wave systematically and gives the consistent result. 

For the incoherent and independent sources, 
The incoherent effect is introduced by averaging the phase angle 
and the interference terms disappear. 

Note that in our method of the double slit state in Fock space, 
single slit states $|A;k>$ and $|B;k>$ are not 
independent but have the non-zero inner product $<B;k|A;k>\neq 0$, 
which tends to the norm of the state A: $<A;k|A;k>$ for zero slit distance 
$(d\rightarrow 0)$ and zero for large slit distance $(d\rightarrow \infty)$. 
This result is thought to be reasonable. 
Our method is quite different from the one in which  
the incoherent sources are treated to be independent $<B;k|A;k>=0$, 
which do not correspond to the real situation of the 
double slit spherical waves. 

The extension of the double slit experiment in our quantum field theory 
to the quantum electrodynamics is straightforward and the effect 
of the helicity vectors is interesting {\cite{Wolf}}.  
The work will appear in a separate paper. 
Our method also can be applied to the Hanbury Brown and Twiss intensity interferometry 
effect {\cite{Hanbury,Purcell,Paul}} 
and the result will appear in another paper.

\vspace{5mm}
\noindent
{\Large{\bf Acknowledgements}}\\

The authors thank Dr. Kazuyasu Shigemoto for useful comments. 
The authors also thank Prof. Nobuyuki Imoto for his advice on reference books. 

%%%%%%%%%%%%%%%%%%%%%%%%%%%%%%%%%%%%%%%%%%%%%%%%%%%%%%%%%%%%%%%%%%%%%%%%%%%%%%

\end{document}